# High critical fields in $MgB_2$ thin films with various resistivity values

**V. Ferrando[a], C. Tarantini[a], P. Manfrinetti[b], D. Marrè[a], M. Putti[a], A. Tumino[a] and C. Ferdeghini[a]**

[a] INFM-LAMIA, Dipartimento di Fisica, Università di Genova, via Dodecaneso 33, 16146 Genova Italy

[b] INFM-LAMIA, Dipartimento di Chimica e Chimica Industriale, via Dodecaneso 31, 16146 Genova, Italy

**Abstract**. In this paper, we analyze the upper critical field of four $MgB_2$ thin films, with different resistivity (between 5 to 50 $\mu\Omega$cm) and critical temperature (between 29.5 to 38.8 K), measured up to 28 Tesla. In the perpendicular direction the critical fields vary from 13 to 24 T and we can estimate 42-57 T range in other direction. We observe linear temperature dependence even at low temperatures without saturation, in contrast to BCS theory. Considering the multiband nature of the superconductivity in $MgB_2$, we conclude that two different scattering mechanisms influence separately resistivity and critical field. In this framework, resistivity values have been calculated from $H_{c2}(T)$ curves and compared with the measured ones.

## 1. Introduction

Since the discovery of superconductivity in magnesium diboride [1], several unusual properties arising from the presence of two distinct s-wave gaps have been emphasized. These gaps are associated with two different sets of bands [2,3]: two $\pi$-bands are nearly isotropic, and two $\sigma$-bands are nearly two-dimensional. These last ones determine the anisotropy of physical properties. Moreover, due to the different parity of the $\sigma$ and $\pi$ bands, the inter-band impurity scattering is negligible compared with the intra-band ones; thus, $\sigma$ and $\pi$ bands can be considered as different channels conducting in parallel and this can explain some superconducting properties [4-6].

In general, thin films show an important spread in critical temperature and residual resistivity values $\rho_0$. $T_c$ can vary from the optimal value down to 25 K and $\rho_0$ can be in a range from few $\mu\Omega$cm up to hundred of $\mu\Omega$cm, differently from what observed on single crystals, which present optimal critical temperatures and low residual resistivity [7-10]. In respect to the single crystals, the critical field values in thin films are considerably higher (up to tens of Teslas) and the anisotropy of the critical field $\gamma$ is always lower (up to 3.5); $\gamma$ usually decreases when temperature increases, even though in some cases the opposite behavior was also observed [11-19].

The difference between the properties of single crystals and thin films can be ascribed to disorder, surely stronger in thin films, that can both decrease $T_c$ and enhance $H_{c2}$.

In our recent paper [20], we tried to focus the role of disorder in thin films in order to find a relationship among $\rho_0$, $T_c$, and critical fields. For this purpose, samples with very different resistivity were considered. $H_{c2}$ and its anisotropy were studied in the framework of Gurevich model [21], which correlates the critical fields to the diffusivity of each band. We emphasized the presence of two different scattering channels influencing critical fields and resistivity, so as high critical fields can be found in low $\rho_0$ samples. In this paper, in the same set of samples of ref. [20], we deeply apply the model to $H_{c2}(T)$ curves. We will show how the residual resistivity of $\pi$ and $\sigma$ bands can be separately estimated from the upper critical field data. The so obtained values are compared with the measured ones and the general agreement of these data will confirm the goodness of the model we used.

## 2. Sample preparation and characterization

In order to study the influence of disorder on the upper critical field in $MgB_2$, we have measured four different films prepared by standard two-step method [11] on different substrates. The samples, whose thickness is in the range 900-1300 Å, were deposited by pulsed laser ablation following the deposition technique described elsewhere [22]. In the following, they will be referred to as film 1, film 2, film 3 and film 4; their properties are summarized in Table 1. The critical temperature varies from 29.5 K to 38.8 K and normal state resistivity varies between 50 μΩcm and 5 μΩcm. X-ray diffraction measurements indicate a strong *c*-axis orientation of the phase in all the samples a part from film 4, in which a weak (101) reflection (the most intense in powders) seems to be detectable. From ϕ scan measurements, we had clear indications of in plane alignment for film 1 [18], while any evidence of in plane orientation has been detected on the films grown on MgO substrates.

|  | FILM 1 | FILM 2 | FILM 3 | FILM 4 |
|---|---|---|---|---|
| substrate | $Al_2O_3$ c-cut | MgO (111) | MgO (111) | $Al_2O_3$ c-cut |
| c axis, Å | 3.517 | 3.532 | 3.533 | 3.519 |
| $T_C$, K | 29.5 | 32 | 33.9 | 38.8 |
| $\Delta T_C$, K | 2.0 | 1.5 | 1.1 | 1.0 |
| RRR | 1.2 | 1.3 | 1.5 | 2.5 |
| $\rho$(40K), μΩcm | 40 | 50 | 20 | 5 |

**Table 1**. Main properties of the four thin films. The critical temperature value is obtained as the 90% of the normal state resistance and the transition width is calculated between 90 % and 10 % of the normal state resistance. The absolute value of resistivity is with an accuracy of 20% due to the uncertainty in thickness determination. For comparison, the *c*-axis of the bulk is 3.521 Å.

## 3. Normal state resistivity and $H_{c2}$

### 3.1 Normal state resistivity

The high structural disorder and nanostructure generally leads to a higher resistivity and lower residual resistivity ratio in thin films compared with single crystals. At the moment, films with residual resistivity $\rho_0$ of the order of few $\mu\Omega$cm [23] and resistivity curves very similar to those of single crystals are available, and film 4 is one of them.

Table 1 summarizes some data drawn from the resistivity curves. We point out that the values of the resistivity at 40 K and the resistivity slope, $d\rho/dT$(300 K), due to the uncertainty in the film thickness evaluation, have an uncertainty of 20%, but the following discussion is not affected by such indetermination.

In these set of samples we verified [20] that the inter-band scattering rate [27,28] can be responsable for the $T_c$ suppression, but in any case it remains negligible with respect to the intra-band ones. Therefore $\rho_0$ is given by the parallel of $\rho_{0\sigma}$ and $\rho_{0\pi}$, the residual resistivities of $\sigma$ and $\pi$ bands, respectively. To determine which band influences more the conduction, the slope of resistivity curves can be considered and compared with theoretical values [24]. The $d\rho/dT$(300 K) values of Table 1 for the four films are reported in figure 1 as function of $\rho_0$: the two dashed regions in the figure represent the theoretical values of $d\rho_\sigma/dT$ and $d\rho_\pi/dT$ considering the spread of the value of parameters in literature (coupling constant and plasma frequencies). The slopes of resistivity for the films are close to the $d\rho_\pi/dT$ value; only film 2 has an intermediate slope between $d\rho_\sigma/dT$ and $d\rho_\pi/dT$, but however closer to $d\rho_\pi/dT$. This means that, in the films here presented, the $\pi$ conduction seems to prevail and so we can assume $\rho \approx \rho_\pi < \rho_\sigma$. This could be means that disorder is especially effective in the B-planes. As comparison, in the same figure, the slopes at room temperatures for various high purity policrystalline bulks [25,26] are also reported. In this case the derivative are considerably higher indicating that the $\sigma$ bands also contribute to the conduction. Really, the analysis of resistivity data as a tool to extract information on multiband effects in $MgB_2$ has been questioned by Rowell [29]. In his paper, he showed that the calculated resistivity can be overestimated due to grain boundary scattering and poor connectivity between grains that can make the real geometrical factor hard to estimate. Actually, we want to point out that, if we overestimate in such way the resistivity of our films, the real dρ/dT values should be lower and the conclusion about the main conduction of π bands should remain even more valid.

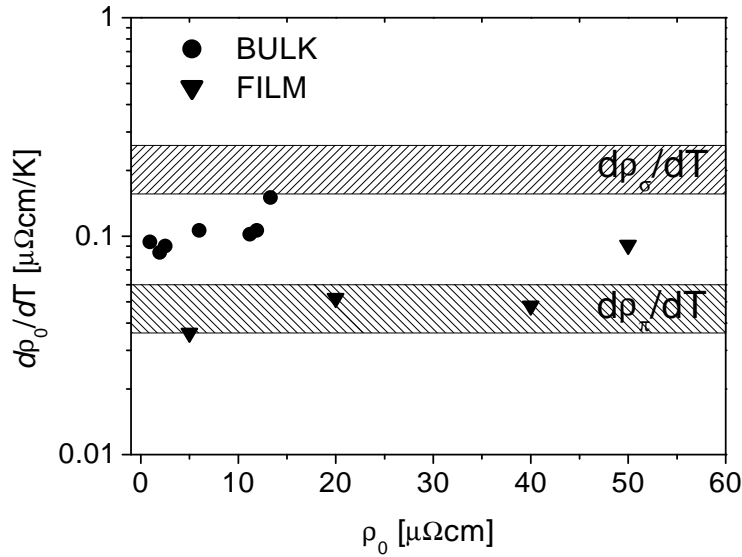

**Figure 1**. $d\rho/dT$(300 K) as a function of $\rho_0$ for the four thin films: the two dashed regions in the figure represent the theoretical values of $d\rho_\sigma/dT$ and $d\rho_\pi/dT$ considering the spread of the value of parameters in literature. As comparison, $d\rho/dT$(300 K) for various bulk are also reported (see text).

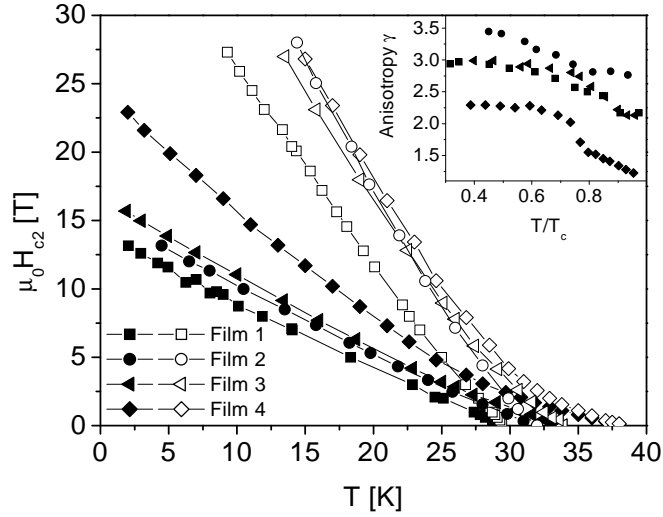

**Figure 2**. Upper critical fields parallel (open symbols) and perpendicular (full symbols) to the basal plane as a function of temperature for the four films. In the inset, anisotropy γ versus $T/T_c$ for the same films.

*3.2. Upper critical field*

Measurements in high magnetic field were performed at GHMFL (Grenoble High Magnetic Field Laboratory) up to 28 T and down to 2 K using a standard four-probe AC technique, in order to evaluate upper critical fields both parallel ($H_{c2}^{\parallel}$) and perpendicular ($H_{c2}^{\perp}$) to the *ab* planes. $H_{c2}^{\parallel}$ and $H_{c2}^{\perp}$ have been estimated for each temperature as the point of the transition in which the resistance is 90% of the normal state value.

Figure 2 shows upper critical field curves as a function of temperature for all the samples. It is worth noting that all $H_{c2}^{\perp}$ (T) curves are parallel despite the great difference in critical temperature and resistivity values of these films. Moreover, they remain linear even at lowest temperature we measured (2 K), so as a reasonable linear extrapolation of $H_{c2}^{\perp}(0)$ is possible obtaining values in the 14-17 T(see Table 2). Film 4 shows a higher $H_{c2}^{\perp}$ (25T): as previously observed, this sample is not completely *c*-oriented and this can induce an overestimation of the smaller critical field ($H_{c2}^{\perp}$), the parallel one being not affected. Therefore we can compare $H_{c2}^{\parallel}$ curves: they show an upward curvature near $T_c$, becoming more evident when the critical temperature approaches the optimal value [12,19], and a linear behavior at low temperature, with similar slope even if $\rho_0$ is very different.

In standard BCS theory, the upper critical field curve should saturate at low temperature and $H_{c2}(0)$ is given by :

$$H_{c2}(0) = 0.69 T_c \left( \frac{dH_{c2}}{dT} \right)$$

where $dH_{c2}/dT$ is proportional to the normal state residual resistivity. $H_{c2}(T)$ curves of MgB$_2$ thin films show a quite different trend. Actually, we do not observe any saturation of $H_{c2}$ at low temperature and BCS extrapolation strongly underestimates the real $H_{c2}(0)$. For example, the BCS zero-temperature extrapolations are $H_{c2}^{\parallel}(0)$=22T and $H_{c2}^{\perp}(0)$=8.75 Tesla for film 1, whereas these values are already reached at 13 and 10 K, respectively. Moreover, in contrast with BCS prediction, our data do not exhibit any $H_{c2}$ dependence on residual resistivity: in the four samples, the critical fields values are similar even though the resistivity values vary from 5 to 50 μΩcm.

In the inset of figure 1 the temperature dependence of anisotropy factor $\gamma = H_{c2}^{\parallel}/H_{c2}^{\perp}$ for all the films is reported. All the curves decrease with increasing temperature. The γ values, at the lowest temperature measured, range between 3 and 3.5 (the maximum value of anisotropy reported for thin films up to now). The low γ observed in film 4, instead, is probably due to the not perfect orientation. Actually, it is not well clarified why thin films show smaller γ values than single crystals: a possible explanation is that disorder makes MgB$_2$ more isotropic.

Since the anisotropy curves of figure 1 seem to saturate at low temperature, we can use the γ values at the lowest temperature to estimate $H_{c2}^{\parallel}(0)$ from the measured $H_{c2}^{\perp}(0)$. They are also reported in Table 2 and vary from 42 to 57 Tesla. Obviously these values can be of great interest for high field application of MgB$_2$.

The two-band nature of superconductivity in MgB$_2$ has to be considered to study upper critical fields. Recently in literature some paper, describing $H_{c2}$ in two bands system both in clean [30] and in dirty limit [21,31,32] apper. In particular, the model proposed by Gurevich [21] for samples in dirty limit takes into account the ratio η between intra-band electronic diffusivities $D_\pi$ and $D_\sigma$, neglecting the inter-band scattering. The upper critical field is determined by the smaller or larger one depending on the temperature range considered. In particular, at low temperature $H_{c2}$ is always dominated by the lowest diffusivity. There are three different conditions marked by γ: if $D_\sigma \ll D_\pi$, γ increases when temperature decreases, while in the opposite case γ decreases by increasing temperature. For $D_\sigma \sim D_\pi$, finally, γ is nearly constant, slightly decreasing as temperature increases; this letter case is similar to BCS one.

|  | FILM 1 | FILM 2 | FILM 3 | FILM 4 |
|---|---|---|---|---|
| $D_\sigma$, m$^2$s$^{-1}$ | 0.48·10$^{-3}$ | 0.48·10$^{-3}$ | 0.46·10$^{-3}$ | 0.37·10$^{-3}$ |
| $D_\pi$, m$^2$s$^{-1}$ | 2.88·10$^{-3}$ | 1.44·10$^{-3}$ | 3.22·10$^{-3}$ | 22.2·10$^{-3}$ |
| η=$D_\pi/D_\sigma$ | 6 | 3 | 7 | 60 |
| ε=$D_\sigma^{(c)}/D_\sigma^{(ab)}$ | 0.11 | 0.08 | 0.11 | 0.19 |
| $\rho_\sigma$, μΩcm | 125 | 125 | 129 | 168 |
| $\rho_\pi$, μΩcm | 14 | 28 | 13 | 2 |
| $\rho_{Hc2}$ μΩcm | 13 | 23 | 12 | 2 |
| ρ measured μΩcm | 40 | 50 | 20 | 5 |
| $H_{c2}(0) \perp$ ab, Tesla | 14.2 | 15.5 | 16.8 | 24.6 |
| γ | 3.0 | 3.5 | 3.0 | 2.3 |
| $H_{c2}(0)$ //ab, Tesla | 42 | 54 | 50 | 57 |

**Table 2**. Some data drawn from resitivity curves and from critical field curves for the four films.

Observing the γ temperature dependences of our films (see figure 1) they seem to indicate that $D_\pi \geq D_\sigma$ and so $\eta = D_\pi/D_\sigma > 1$. This result is the same we obtained by normal state resistivity analysis: in fact, we found $\rho_\pi < \rho_\sigma$ for all the films, that implies $D_\pi > D_\sigma$. In particular, according to what reported in [21], from γ(T) curves it is possible to evaluate not only η, the ratio between the diffusivities of the two bands in the *ab* plane but also the anisotropy of the σ band $\varepsilon = D_\sigma^{(c)}/D_\sigma^{(ab)}$. Being the π band nearly isotropic, its diffusivity has been considered the same in the two directions. From the ratio between $H_{c2}$ parallel and perpendicular to the c axis in fact (see eq.(36), (56) and (57) in ref.[21]), the following expressions for the anisotropy at T=0 and T=$T_c$ can be obtained:

$$\gamma(T_c) = \frac{(a_1 + a_2\eta)}{(a_1\varepsilon^{1/2} + a_2\eta)}$$

$$\gamma(0) = \frac{\exp\left(\frac{g(\pi/2) - g(0)}{2}\right)}{\varepsilon^{1/4}}$$

with

$$g(\theta) = \left(\frac{\lambda_0^2}{w^2} + \ln^2\frac{D_\pi}{D_\sigma(\theta)} + \frac{2\lambda_-}{w}\ln\frac{D_\pi}{D_\sigma(\theta)}\right)^{1/2} - \frac{\lambda_0}{w}$$

$a_1$, $a_2$, $\lambda_0$, $\lambda_-$ and *w* are defined in ref.[21] and depend from coupling constants; their values are 1.93, 0.07, 0.564, 0.525 and 0.22 respectively for $MgB_2$. According to the field orientation, $D_\sigma(0) = D_\sigma^{(ab)}$ and $D_\sigma(\pi/2) = (D_\sigma^{(ab)} D_\sigma^{(c)})^{1/2}$. Therefore also γ($T_c$) and γ(0) are functions of η and ε and we can determine them analysing anisotropy data. In Table 2, η and ε values for the four films are reported, together with the $D_\sigma$ which can be now estimated from the measured $H_{c2}^\perp(0)$. First of all, we can notice that in our film, in which η > 1, $\varepsilon^{-1/2} = (D_\sigma^{(ab)}/D_\sigma^{(c)})^{1/2} \approx \gamma(0)$; this means that anisotropy of critical fields at zero-temperature is mainly determined by anisotropy of σ band. From this model, if η < 1, there is a different relation between ε and anisotropy; in fact, in this case, it would obtain $\varepsilon^{-1/2} \approx \gamma(T_c)$. Moreover, the obtained values for $D_\sigma$ are similar for film 1, 2 and 3 (around $0.48 \cdot 10^{-3}$ m$^2$s$^{-1}$) and slightly lower only for film 4, which presents higher $H_{c2}^\perp$ value. Using these $D_\sigma$ and η values, the π diffusivity $D_\pi$ can also be obtained (see Table 2). Finally the resistivities associated to the σ and π bands, $\rho_\sigma$ and $\rho_\pi$, can be inferred by

$$\frac{1}{\rho_{\sigma,\pi}} = e^2 N_{\sigma,\pi} D_{\sigma,\pi}$$

The calculated $\rho_\sigma$ and $\rho_\pi$ are reported again in Table 2. $\rho_\sigma$ is very similar in all the films and ranges between 125 and 168 μΩcm, while $\rho_\pi$ varies of more than one order of magnitude, ranging from around 2 to 28 μΩcm. This is an important result: the two band nature of $MgB_2$ leads to two different scattering mechanisms which influence separately critical fields and resistivity. In particular, in our case, the high σ-bands resistivity induces the high critical fields, while resistivity is essentially determined by π-band. If interband scattering is negligible with respect to the intra-band ones as in this case, the overall resistivity can be calculated from the parallel between σ and π ones. The results of this calculation for all the samples are presented again in Table 2 as $\rho_{Hc2}$. It must be noted that these values, totally obtained from $H_{c2}$ measurements, are in reasonable agreement with the measured ones, differing of about a factor two from the measured resistivities. This fact, together with the low Δρ of our samples, seems to confirm that our resistivity data are not strongly influenced by grain boundary scattering and lack of connectivity between grains. The slight discrepancy between the measured resistivities and values obtained from $H_{c2}$ measurement can be partially ascribed to a difference between the theoretical and experimental values of off-diagonal coupling constants, as hypothesized in ref. [33]. A small variation of these parameters could slightly increses the values of $\rho_{H_{c2}}$. Finally, what is peculiar in our films is that the σ band is much more dirty than π band. This could be due to a disorder in the B-planes formed during the deposition process, that could be poorly recovered during annealing in Mg atmosphere for the phase crystallization.

## 4. Conclusions

In order to clarify the role of disorder in magnesium diboride, four thin films with different values of resistivity and critical temperature have been studied. We suggest that the Tc suppression is determined by the inter-band impurity scattering. Very high $H_{c2}$ values, up to 24 T in perpendicular direction and up to 57 T in the parallel orientation have been found in samples with low $\rho_0$

Upper critical fields and anisotropy have been studied using the model proposed by Gurevich, taking into account the two band nature of superconductivity in $MgB_2$. We evidenced how the scattering mechanisms determining critical field and resistivity can be different. This analysis explains why films with resistivities varing by one order of magnitude can show similar critical fields. Moreover, we were able to calculate a resistivity value directly from $H_{c2}$ data: the calculated values resulted to be in good agreement with the measured ones.

This work was supported by the European Community through "Access to Research Infrastructure action of the Improving Human Potential Programme".